# Dynamical mass generation via space compactification in graphene

A. D. Alhaidari[1], A. Jellal[1,2,3], E. B. Choubabi[3] and H. Bahlouli[4,1]

[1]*Saudi Center for Theoretical Physics, Jeddah, Saudi Arabia*
[2]*Physics Department, College of Science, King Faisal University, PO Box 380, Al-Ahsaa 31982, Saudi Arabia*
[3]*Theoretical Physics Group, Faculty of Sciences, Chouaib Doukkali University, PO Box 20, 24000 El Jadida, Morocco*
[4] *Physics Department, King Fahd University of Petroleum & Minerals, Dhahran 31261, Saudi Arabia*

Fermions in a graphene sheet behave like massless particles. We show that by folding the sheet into a tube they acquire non-zero effective mass as they move along the tube axis. That is, changing the space topology of graphene from 2D to 1D (space compactification) changes the 2D massless problem into an effective massive 1D problem. The size of the resulting mass spectrum depends on the quantized azimuthal frequency and its line spacing is proportional to the inverse of the tube diameter.

About six years ago, the possibility to isolate and investigate graphene, individual one-atom-thick layers of graphite, has been demonstrated using micromechanical cleavage methods [1-3]. Graphene is the name given to a perfect infinite single sheet of carbon bonded atoms in a honeycomb lattice. It was shown that the dynamics of low energy charge carriers in graphene may be described to a high degree of accuracy by a massless Dirac equation for quasi-particles having a linear energy dispersion near the two edges of the hexagonal Brillouin zone (also called Dirac points). This linear energy dispersion resulted in special intriguing electronic transport properties and attracted wide theoretical attention [4] due to its potential applications [1-4]. The 2D structure of graphene has been used to describe properties of many carbon based materials such as graphite (a large number of graphene sheets), nanotubes (nanometer-sized cylinders made of rolled up graphene sheets), fullerenes (spherically shaped graphene sheets) and ribbons (strips of graphene sheets) [5].

In the past, the study of relativistic particles has been the exclusive domain of high-energy and particle physics. In graphene, nonetheless, the linear electronic band dispersion near the Dirac points gave rise to charge carriers (electrons or holes) that propagate as if they were massless fermions with speeds of the order of $10^6$ m/s rather than the speed of light $3\times10^8$ m/s. Hence, charge carriers in this structure should be described by the massless Dirac equation rather than the usual Schrodinger equation. The physics of relativistic electrons is thus now experimentally accessible in graphene-based solid-state devices, whose behavior differs drastically from that of similar devices fabricated with usual semiconductors. Consequently, new unexpected phenomena have been observed while other phenomena that were well-understood in common semi-conductors, such as the quantum Hall effect and weak-localization, exhibited surprising behavior in graphene. Thus, graphene devices enabled the study of relativistic dynamics in controllable nano-electronic circuits (relativistic electrons on-a-chip) and their behavior probes our most basic understanding of electronic processes in solids. It also allowed for the observation of some subtle effects, previously accessible only to high energy physics, such as Klein tunneling and vacuum breakdown.

From the electronic applications point of view, graphene-based materials with gap are very desirable. This lead to substantial theoretical and experimental efforts to generate and control the energy gap in graphene based devices. One approach is based on quantum confinement as in quantum dots and nanoribbons, where it was shown that the energy gap value increases with decreasing nanoribbon width [6]. Another approach used spin-orbit coupling and Rashba interaction [7] and a third approach was based on interlayer coupling [8] to mention only few. In the present work, we look at the energy gap in graphene as inertia or mass generated through space compactification of a higher dimensional system (e.g., graphene sheet rolled into a tube). Dynamical mass generation via space compactification (dimensional reduction) was of interest to high energy physicists a long time ago [9]. The existence of an extra space-like dimension is assumed which, due to quantum corrections, contributes a mass to an originally massless field. In our case, however, we will show that effective mass generation comes naturally from rolling the graphene sheet into a tube; changing the space topology from 2D to an effective 1D. Technically, the tube is described by an effective 1D Dirac Hamiltonian with a constant pseudo-scalar term in the interaction. Moreover, we argue that the size of the resulting mass spectrum depends on the quantized azimuthal frequency and its line spacing is proportional to the inverse of the tube diameter.

The natural coordinates for the 2D graphene tube, of radius $R$, are the cylindrical coordinates: $z \in [-\infty, +\infty]$ and $\theta \in [0, 2\pi]$. The time-dependent Dirac equation (in the units $\hbar = c = 1$) that describes the behavior of the massless fermions on the surface of the tube is

$$i\frac{\partial}{\partial t}\Psi = \left(-i\vec{\alpha}\cdot\vec{\nabla} + e\vec{\alpha}\cdot\vec{A} + eA_0\right)\Psi, \tag{1}$$

where $\vec{\alpha} = (\sigma_\theta, \sigma_z)$ are two Hermitian matrices such that $\sigma_\theta^2 = \sigma_z^2 = 1$ and $\sigma_\theta \sigma_z = -\sigma_z \sigma_\theta$. A natural choice of minimum representation would be in terms of the Pauli matrices as follows: $(\sigma_\theta, \sigma_z) = (\sigma_1, \sigma_2)$. Other possible representations exist but are equivalent to this one. The three-vector potential in 2+1 dimensions is $A_\mu = (A_0, \vec{A})$. Assuming time-independent potentials, the spinor wavefunction could be written as $\Psi(t, z, \theta) = e^{-i\varepsilon t}\psi(z, \theta)$, where $\varepsilon$ is the system energy. Thus, the time independent Dirac equation becomes $(H - \varepsilon)\psi = 0$, where $H = H_0 + \sigma_1 H_\theta + \sigma_2 H_z$, and $H_0 = eA_0$, $H_\theta = -\frac{i}{R}\partial_\theta + eA_\theta$, $H_z = -i\partial_z + eA_z$. $A_0(z)$ is the potential function of interest and for our case we take $A_z = A_\theta = 0$. Thus, the explicit matrix form of the Dirac equation reads as follows

$$\begin{pmatrix} eA_0 - \varepsilon & -\partial_z - \frac{i}{R}\partial_\theta \\ +\partial_z - \frac{i}{R}\partial_\theta & eA_0 - \varepsilon \end{pmatrix}\begin{pmatrix} \psi_+ \\ \psi_- \end{pmatrix} = 0. \tag{2}$$

where $\psi_+$ and $\psi_-$ are the upper and lower spinor components, respectively.

If we write $\psi_\pm(z, \theta) = \phi_\pm(z)\chi_\pm(\theta)$ then the above equation becomes completely separable if $\chi_+ = \chi_- = \chi$ and $d\chi/d\theta = i\mu$, where $\mu$ is a real constant. Therefore, we can write $\chi(\theta) = e^{i\mu\theta}$ where $|\mu|/R$ stands for the fermions azimuthal frequency $\omega$. The boundary condition $\psi(z, \theta) = \psi(z, \theta + 2\pi)$ dictates that $\mu$ is an integer. That is, $\mu = 0, \pm 1, \pm 2, .., \pm N$, where $N$ is the integral part of $\omega R$. Eq. (2) then reduces to

$$\begin{pmatrix} eA_0 - \varepsilon & -\frac{d}{dz} + \frac{\mu}{R} \\ +\frac{d}{dz} + \frac{\mu}{R} & eA_0 - \varepsilon \end{pmatrix} \begin{pmatrix} \phi_+ \\ \phi_- \end{pmatrix} = 0, \tag{3}$$

giving the relation between the two spinor components:

$$\phi_\mp(z) = \frac{1}{\varepsilon - eA_0(z)}\left(\pm\frac{d}{dz} + \frac{\mu}{R}\right)\phi_\pm(z). \tag{4a}$$

If the potential $A_0(z)$ is constant over a given interval of the $z$-axis, then we also obtain the second order differential equation:

$$\left[\frac{d^2}{dz^2} + (\varepsilon - V)^2 - \left(\frac{\mu}{R}\right)^2\right]\phi_\pm(z) = 0, \tag{4b}$$

where we have defined the constant $V = eA_0$. This equation has the following general solution on the same interval

$$\phi_\pm(z) = Ae^{kz} + Be^{-kz}, \tag{5}$$

where $A$ and $B$ are complex constants that depend on the energy and the physical parameters, and $k = \sqrt{(\mu/R)^2 - (\varepsilon - V)^2}$ is the wave number. Hence, we obtain oscillatory solutions (continuum scattering states) if $|\varepsilon - V| > |\mu/R|$ and exponential solutions (bound states) if $|\varepsilon - V| < |\mu/R|$. It is very interesting to note that the dispersion relation associated with the massive 1D Dirac equation for the same constant vector potential $V$ is $k = \sqrt{m^2 - (\varepsilon - V)^2}$ [10]. Thus, the massless 2D problem seems to be analogous to the corresponding massive 1D problem with mass $m = |\mu/R|$. Below, we prove that there is indeed an exact equivalence between the two problems, the proof is provided both analytically and numerically. This feature is very important and could be used to discuss the zero mode energy and gap creation in graphene.

Now, Eq. (4a) and Eq. (4b) lead to the following two alternative solutions

$$\psi(z,\theta) = Ae^{i\mu\theta + kz}\begin{pmatrix} 1 \\ \gamma_+ \end{pmatrix} + Be^{i\mu\theta - kz}\begin{pmatrix} 1 \\ \gamma_- \end{pmatrix}, \tag{6a}$$

$$\psi(z,\theta) = Ae^{i\mu\theta + kz}\begin{pmatrix} \gamma_- \\ 1 \end{pmatrix} + Be^{i\mu\theta - kz}\begin{pmatrix} \gamma_+ \\ 1 \end{pmatrix}, \tag{6b}$$

on the given $z$-axis interval where $\gamma_\pm = \frac{\pm k + \mu/R}{\varepsilon - V}$. Therefore, we end up with two possible independent solutions. That is, there is an ambiguity resulting from the choice of sign (top or bottom) in Eq. (4a) and Eq. (4b) that should be resolved. One way to settle this ambiguity is to resort to the above-mentioned analogy with the 1D massive case. This analogy implies that if $\varepsilon > V$ then Eq. (4b) with the top sign gives the upper component, $\phi_+(z)$, that belongs to the positive energy subspace and Eq. (4a) with the top sign gives the lower component, $\phi_-(z)$, that belongs also to the same positive energy subspace. On the other hand, if $\varepsilon < V$ then Eq. (4b) and Eq. (4a) with the bottom signs give $\phi_-(z)$ and $\phi_+(z)$, respectively. These latter solutions belong to the negative energy subspace. Another way to resolve the ambiguity issue is to investigate the action of the raising/lowering operator $\left(\pm\frac{d}{dz} + \frac{\mu}{R}\right)$ on the state $e^{i|k|z}$, which will lead to the same conclusion. Hence, for $\varepsilon > V$, (6a) is the solution that belongs to the positive energy subspace, whereas if $\varepsilon < V$ then (6b) is the solution, which belongs to the negative energy subspace.

To gain more insight into the analogy between the present 2D massless system and the corresponding massive 1D problem, we consider the simple square barrier potential of height $V$ and width $a$. This system could be realized experimentally by inserting the long graphene tube into a short cylindrical capacitor of length $a$ and charging it to a potential $V$. Our strategy is to obtain the scattering solution in this setting then compare these results to the solution of the problem associated with the massive 1D Dirac equation:

$$\begin{pmatrix} m + V(x) - \varepsilon & -\frac{d}{dx} \\ +\frac{d}{dx} & -m + V(x) - \varepsilon \end{pmatrix} \begin{pmatrix} \phi_+(x) \\ \phi_-(x) \end{pmatrix} = 0, \qquad (7)$$

where $V(x)$ is the same square barrier potential and the mass $m$ is equal to $|\mu/R|$ (see Fig. 1). Scattering in the 2D problem occurs for energies $\varepsilon > |\mu/R|$ and the associated boundary conditions give the following eigen-spinors outside the barrier:

$$z < 0: \qquad \phi(z) = \begin{pmatrix} 1 \\ \beta \end{pmatrix} e^{iqz} + R \begin{pmatrix} 1 \\ \beta^* \end{pmatrix} e^{-iqz}, \qquad (8a)$$

$$z > a: \qquad \phi(z) = T \begin{pmatrix} 1 \\ \beta \end{pmatrix} e^{iqz}, \qquad (8b)$$

where $q = \sqrt{\varepsilon^2 - (\mu/R)^2}$ is the wave number and $\beta = \left(iq - \frac{\mu}{R}\right)/\varepsilon$. These represent a normalized flux of particles with energy $\varepsilon$ incident from one end of the tube that gets partially transmitted to the other end with an amplitude $T(\varepsilon)$ and partially reflected with an amplitude $R(\varepsilon)$. These amplitudes depend not only on the energy but also on the other physical parameters $\{\mu, R, V, a\}$. Unitarity of the problem results in the current (particle flux) conservation equation, $|T|^2 + |R|^2 = 1$. Note that for positive (negative) energy, $e^{\pm iqz}$ is a wave traveling in the $\pm z$ ($\mp z$) direction, respectively. The solution inside the barrier ($0 < z < a$) is

$$\varepsilon > V: \qquad \phi(z) = Ae^{kz} \begin{pmatrix} 1 \\ \gamma_- \end{pmatrix} + Be^{-kz} \begin{pmatrix} 1 \\ \gamma_+ \end{pmatrix}, \qquad (9a)$$

$$\varepsilon < V: \qquad \phi(z) = Ae^{-kz} \begin{pmatrix} \gamma_- \\ 1 \end{pmatrix} + Be^{kz} \begin{pmatrix} \gamma_+ \\ 1 \end{pmatrix}. \qquad (9b)$$

Matching the wavefunction (9a) or (9b) with (8a) at $z = 0$ and with (8b) at $z = a$ results in a set of four equations that are solved for the four unknowns $A$, $B$, $R$, and $T$.

For a given choice of physical parameters, Fig. 2 shows a comparison between the result of our calculation of the transmission coefficient for the massless fermions in the graphene tube associated with Eq. (3) and for the corresponding massive 1D problem associated with the Dirac equation (7). The physical parameters of the two problems are related by $m = |\mu|/R$. The exact match implies that by folding the graphene sheet into a tube, fermions acquire non-zero effective mass as they move along the tube axis. That is, changing the space topology of graphene from 2D to 1D (space compactification or dimensional reduction) changes the 2D massless problem into an effective massive 1D problem. Moreover, the size $N$ of the resulting mass spectrum $\{n/R\}_{n=0}^{N}$ depends on the azimuthal frequency $\omega$ and its line separation, $\Delta n/R$, is proportional to the inverse of the tube diameter. From a mathematical point of view, the exact match between the two problems could be explained as follows. There exists a unitary transformation that takes the Hamiltonian operator of one problem into the other. That is, we can find a 2×2 unitary matrix that maps Eq. (3) into Eq. (7). One can show that this transformation matrix could be written as $e^{\pm i(\pi/4)\sigma_2} = \frac{1}{\sqrt{2}}\begin{pmatrix} 1 & \pm 1 \\ \mp 1 & 1 \end{pmatrix}$, where ± is the sign of $\mu$. Physically, however, the similarity between the two problems is subtle and highly non-trivial. For example, the massless Dirac Hamiltonian with vector, $V(x)$, scalar, $S(x)$, and pseudo-scalar, $W(x)$, potentials

$$H = \begin{pmatrix} V(x)+S(x) & -\frac{d}{dx}+W(x) \\ +\frac{d}{dx}+W(x) & V(x)-S(x) \end{pmatrix}, \qquad (10)$$

has been considered extensively in the literature for various problems either in one dimension [11] or, after variable separation, in higher dimensions [12]. However, the pseudo-scalar term in the Hamiltonian of Eq. (3) is a global term (i.e., it is constant over the whole configuration space). This leads to the conclusion that the dynamical mass generated by space compactification of the graphene sheet has a pseudo-scalar character rather than the usual scalar one. Recall that this method of mass generation has been utilized exclusively in high energy physics, supergravity, string theory and related fields [9]. To the best of our knowledge, the present work constitutes the first successful application of this method in condensed matter physics. Another example of space compactification is found in a system consisting of a stack of graphene sheets with coupling between the layers making the massless 3D problem equivalent to an effective massive 2D problem [8].

**Acknowledgments**:
The generous support provided by the Saudi Center for Theoretical Physics (SCTP) is highly appreciated by all Authors. We are also grateful to Les Résidences d'Ifrane (Ifrane, Morocco) for hosting the meeting where part of this work was conducted. AJ acknowledges partial support by King Faisal University, and HB acknowledges partial support by King Fahd University of Petroleum & Minerals.

**Figures Captions:**

**Fig. 1:** The square potential barrier with a height larger than $2|\mu|/R$. The positive/negative energy oscillatory solution is represented by the light/dark grey area. The white area represents the bound state exponential solution. Fermions and anti-fermions are represented by the outlined arrows (⇨) and solid arrows (➔), respectively.

**Fig. 2:** The transmission coefficient as a function of energy for the massless problem associated with the graphene tube (solid curve) and for the corresponding massive 1D problem (dashed curve). In the units $\hbar = c = 1$, we took the barrier height $V = 4$ and width $a = 3$. The physical parameters of the two problems are related by $m = |\mu|/R$ which is set to unity. The figure demonstrates an exact match and shows transmission resonances in the Klein energy zone ($1 < \varepsilon < 3$).

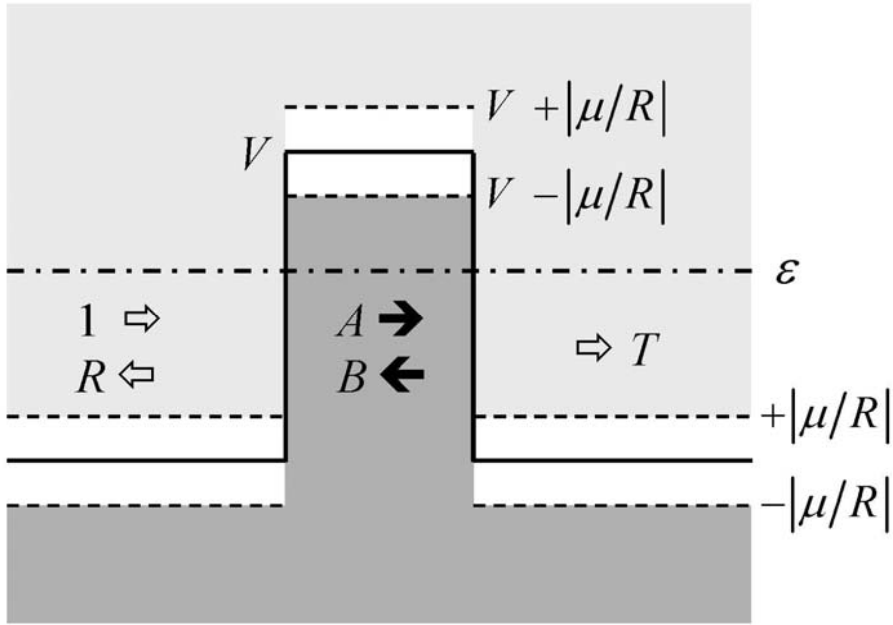

**Fig. 1**

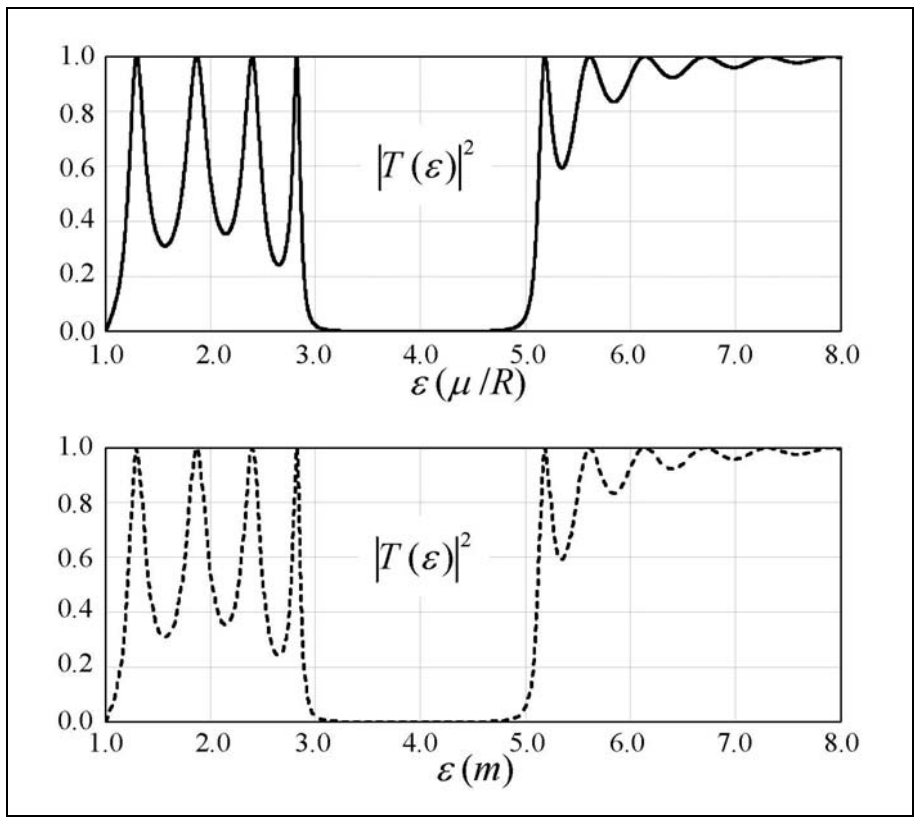

**Fig. 2**